\documentclass[twocolumn,showpacs,prb,ams]{revtex4}
\usepackage{graphicx}
\usepackage{graphics}
\usepackage{dcolumn}
\usepackage{bm}
\usepackage{amsmath}
\begin{document}
\title{Magnetism near the metal-insulator transition in two-dimensional
electron systems: the role of interaction and disorder}

\author{S. M. Fazeli and K. Esfarjani}
\affiliation{Department of Physics, Sharif University of Technology,
Tehran 11365-9161, Iran\\}
\author{B. Tanatar}
\affiliation{Department of Physics, Bilkent University, 06800 Bilkent,
Ankara, Turkey}
\date{\today}
\begin{abstract}
Recent thermodynamic measurements on two-dimensional (2D) electron
systems have found diverging behavior in the 
magnetic susceptibility and appearance of ferromagnetism with decreasing
electron density. The critical densities for these phenomena
coincide with the metal-insulator transition recorded in
transport measurements. Based on density functional calculations within
the local spin-density approximation, we have investigated the
compressibility and magnetic susceptibility of a 2D electron gas
in the presence of remote impurities.
A correlation between the minimum in the inverse capacitance 
($\partial\mu /\partial n$) and the maximum of magnetization and 
magnetic susceptibility is found. Based on values we obtain for the inverse participation 
ratio, this seems to be also the MIT point.

\end{abstract}

\pacs{71.10.-w, 71.10.Ca, 71.30.+h}
\maketitle

{\it Introduction}--There has been a large amount of experimental
and theoretical activity
in recent years to understand the ground state properties of
two-dimensional (2D) electron systems\cite{abrahams}.
Most notably, the observation of a metal-insulator
transition (MIT) in these systems provides a major motivation
to study the various physical properties. Most experiments perform
transport measurements obtaining resistivity or conductivity as a
function of temperature at varying electron density to deduce the
metallic or insulating phases.
In contrast, Ilani {\it et al}.\cite{ilani} used the capacitance
technique, a thermodynamic measurement, to measure the compressibility
finding that at high densities the results for the inverse capacitance
($\partial\mu /\partial n$) agree with Hartree-Fock or mean-field theories, 
whereas below the critical density the system becomes very inhomogeneous 
and $\partial\mu /\partial n$ goes up as the density is lowered. 
Similar observations were also made by Dultz and
Jiang\cite{dultz} on a 2D hole system who noted that the inverse
compressibility is minimum at the same density where the
MIT occurs.

Recent experiments at in-plane magnetic field concentrated on the
spin susceptibility, Lande $g$-factor, and
effective mass of the 2D electron systems present in
Si-MOSFETS and GaAs quantum-well
structures\cite{pudalov,tutuc,zhu,pan,prus,shashkin05}.
In particular, Shashkin {\it et al}.\cite{shashkin1,shashkin2}
reported a sharp increase of the effective
mass near the critical density at which the system starts to show
deviations from the metallic behavior. On the other hand, Pudalov {\it
et al}.\cite{pudalov} have found only moderate enhancement of the spin
susceptibility and effective mass in their samples.
Thermodynamic measurements of magnetization of a dilute 2D electron
system were reported by Prus {\it et al}.\cite{prus} and Shashkin
{\it et al}.\cite{shashkin05} Both experiments found large
enhancement of the spin susceptibility $\chi_s$ over its Pauli
value.
Whereas the measurements of Prus {\it et al}.\cite{prus} found
no indication toward a ferromagnetic instability, Shashkin
{\it et al}.\cite{shashkin05} observed diverging behavior in
$\chi_s$ at a critical density coinciding with the MIT
density obtained from transport measurements.

On the theoretical side, calculation of compressibility for a 2D
system of electrons in the presence of disorder predicts the observed
behavior of upturn and divergence\cite{sv_at}.
Shi and Xie\cite{xie} performed density functional
calculations based on the (unpolarized) local density approximation 
developed by Tanatar and Ceperley \cite{tc} within the Thomas-Fermi 
(TF) 
theory, and found similar results for the compressibility. They
also identified the MIT point with the percolation transition point
in this system. This theory was further strengthened by Das Sarma
{\it et al}.\cite{dassarma} who measured the critical exponent for the
conductivity and found it in agreement with that proposed by
percolation theory.

In this work, we investigate the spatial distribution of carrier
density and magnetization, as well as the compressibility and
spin susceptibility of a 2D electron system using the local 
spin-density approximation (LSDA) both at the
TF and Kohn-Sham (KS) levels.
Localization properties are characterized by the inverse participation 
ratio (IPR). The functional we use is the
one constructed by Attacallite {\it et al}.\cite{moroni} from
the very recent quantum Monte Carlo calculations for correlation energy
appropriate for uniform systems. An important feature of these 
simulations is that a transition to a ferromagnetic phase at low 
densities is built in the functional.
A disorder potential due to remote impurities is included to make
the calculation realistic. The density distribution of the system
shows high and low density regions as reported earlier by Shi and
Xie\cite{xie} which supports the idea of percolation transition.
We also find correlated behavior of inverse capacitance minimum
and spin susceptibility maximum, which gives thermodynamic
evidence for the appearence of a phase transition. Thus, a unified
picture for the thermodynamic behavior of a dilute system of 2D
electrons in the presence of disorder emerges from our
calculations.

{\it Theory}--We consider a 2D electron system interacting via the
long range Coulomb interaction $V_q=2\pi e^2/(\epsilon q)$ where
$\epsilon$ is the
background dielectric constant. The system is characterized by the
dimensionless interaction strength $r_s=1/(\pi n {a_B^{\ast}}^2)^{1/2}$,
where $n$ is the 2D electron density and $a_B^{\ast}=
\hbar^2\epsilon/(m^* e^2)$ is the effective Bohr radius defined
in terms of the band mass $m^*$ of electrons in the bulk semiconductor
structure.

Within the spin-density functional theory the total energy of an
$N$-electron interacting system in a local external potential $V_{\rm
ext}({\bf r})$ is a unique functional of spin-dependent densities 
$n_\uparrow({\bf r})$ and $n_\downarrow({\bf r})$. 
The total energy functional can be expressed as
\begin{equation}
\begin{split}
E_{\rm Total}&[n_\uparrow,n_\downarrow]=
E_{T}[n_\uparrow,n_\downarrow]+
E_{H}[n_\uparrow,n_\downarrow] \nonumber\\
&+
E_{x}[n_\uparrow,n_\downarrow]+
E_{c}[n_\uparrow,n_\downarrow]+
E_{\rm ext}[n_\uparrow,n_\downarrow]\, .
\end{split}
\end{equation}
We approximate the kinetic energy functional by the
Thomas-Fermi-Weiz{\"a}cker (TFW) form given by\cite{zaremba}
\begin{equation}
E_T[n_\uparrow,n_\downarrow]=\sum_\sigma\int d{\bf r}\,\left[\pi
n_\sigma^2({\bf r})+\frac{1}{8}\frac{|\nabla n_\sigma({\bf r})|^2}{n_\sigma({\bf
r})}\right]\, .
\end{equation}
The direct Coulomb energy is given by
\begin{equation}
E_H[n_\uparrow,n_\downarrow]=\frac{e^2}{2}\,\int d^2{\bf r}\,d^2{\bf
r'}\,\frac{n({\bf r})\,n({\bf r'})}{|{\bf r-r'}|}
\end{equation}
where $n({\bf r})=n_\uparrow({\bf r})+n_\downarrow({\bf r})$ is the
total density.
The local spin-density approximation proposed by Attaccalite
{\it et al}.\cite{moroni}
is used to calculate the exchange and correlation potentials.

The disorder studied in this work comes from a random distribution
of charged impurities with density $n_i$ and at a setback distance 
$d$ from the electron
layer. The energy functional due to the external potential is
\begin{equation}
E_{\rm ext}[n]=\int d^2{\rm r}\,V_{\rm ext}({\bf
r})\,n({\bf r})
\end{equation}
where the external potential, due to remote impurities,
is given by
\begin{equation}
V_{\rm ext}({\bf r})=-\sum_i \frac{Ze^2/\epsilon}
{[({\bf r-r_i})^2+d^2]^{1/2}}\, .
\end{equation}

{\it Method}--
The spin-densities $n_\uparrow({\bf r})$ and $n_\downarrow({\bf r})$
that extremize the total energy functional can be obtained by
annealing from a Monte Carlo (MC) simulation. A sufficiently high
temperature is first chosen and a Metropolis MC
run is performed long enough to reach thermodynamic equilibrium. Then
the temperature is reduced and the run is repeated.
This is continued until the final temperature is sufficiently low so
that very little energy fluctuations occur during the last run.
This simulation is done in order to reach the
global minimum of the energy landscape. Once one is near the bottom
of the valley, eventually a steepest-descent algorithm is applied
to reach the minimum energy structure faster.
The areal integral in the energy functional is approximated by a
discrete sum: basically the density and potentials are discretized
on a 44 by 44 mesh whose size is one effective Bohr radius being
equal to $100$\,\AA\ in GaAs samples. The long-range Coulomb
potential is calculated using the Ewald sum method.

The results from the
TFW kinetic energy functional were also compared to the solution of
the KS equations where the kinetic energy is calculated exactly.
We note that a similar Hartree-Fock calculation was recently
performed on a Mott insulator in the half-filled
limit\cite{trivedi} where the ground state is antiferromagnetic.
We believe that the exchange-correlation potential we are using is
more realistic, and contains the correct physics, namely that at
very low densities, before the Wigner crystallization, the clean system
becomes ferromagnetic.

{\it Results and Discussion}--For a given distribution of the
impurities, the total energy is computed as a function of density.
The obtained curve was then fitted with 3 best analytical curves
possible and then derivatives were taken analytically in order to
reduce fluctuations. The inverse compressibility is obtained as
$\kappa^{-1}= n^2 \partial\mu /\partial n$. There is some
variations in the results at higher densities, but as Fig.\,1
shows, both the minimum and the zeros of $\partial\mu /\partial n$ are
identical with reasonably good accuracy.
\begin{figure}[h]
\begin{center}
\includegraphics[width=8 cm]{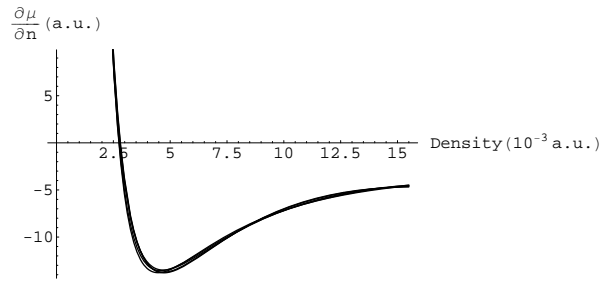}
\caption{(Color online) The derivative of the chemical potential versus density,
expressed in units of the effective Bohr radius, obtained with the TFW functional.
The disorder potential corresponds to a spacer layer of $d=3$\,a.u.
and a 2D impurity concentration of $n_i=2.6\times 10^{-3}$\,a.u. both
expressed in units of the effective Bohr radius.}
\end{center}
\label{kappa}
\end{figure}

The electron gas was found to have a spin polarization
$\xi(r)=[n_{\uparrow}(r)-n_{\downarrow}(r)]/n(r)$ reaching
a maximum of about 0.05 at $B=0$ in the low density regions.
At lower densities below the minimum in the curve for
$\partial \mu /\partial n$ the electron gas separates into
electron puddles where the disorder potential is lowest. Above that
critical density there seems to be percolation as noted by Shi and
Xie.\cite{xie}
The magnetization per electron ($\xi(r)$), as can be observed in
Fig.\,2, was found to be largest at the critical point
(minimum of $\partial \mu /\partial n$) and it was found that both $\xi$ and the
susceptibility are highest in the low density regions where the depth of the effective potential($V_{\rm eff}=V_{\rm ext}+V_{\rm Hartree}+V_{\rm xc}$) is most shallow.
\begin{figure}[h]
\begin{center}
\includegraphics[width=6 cm]{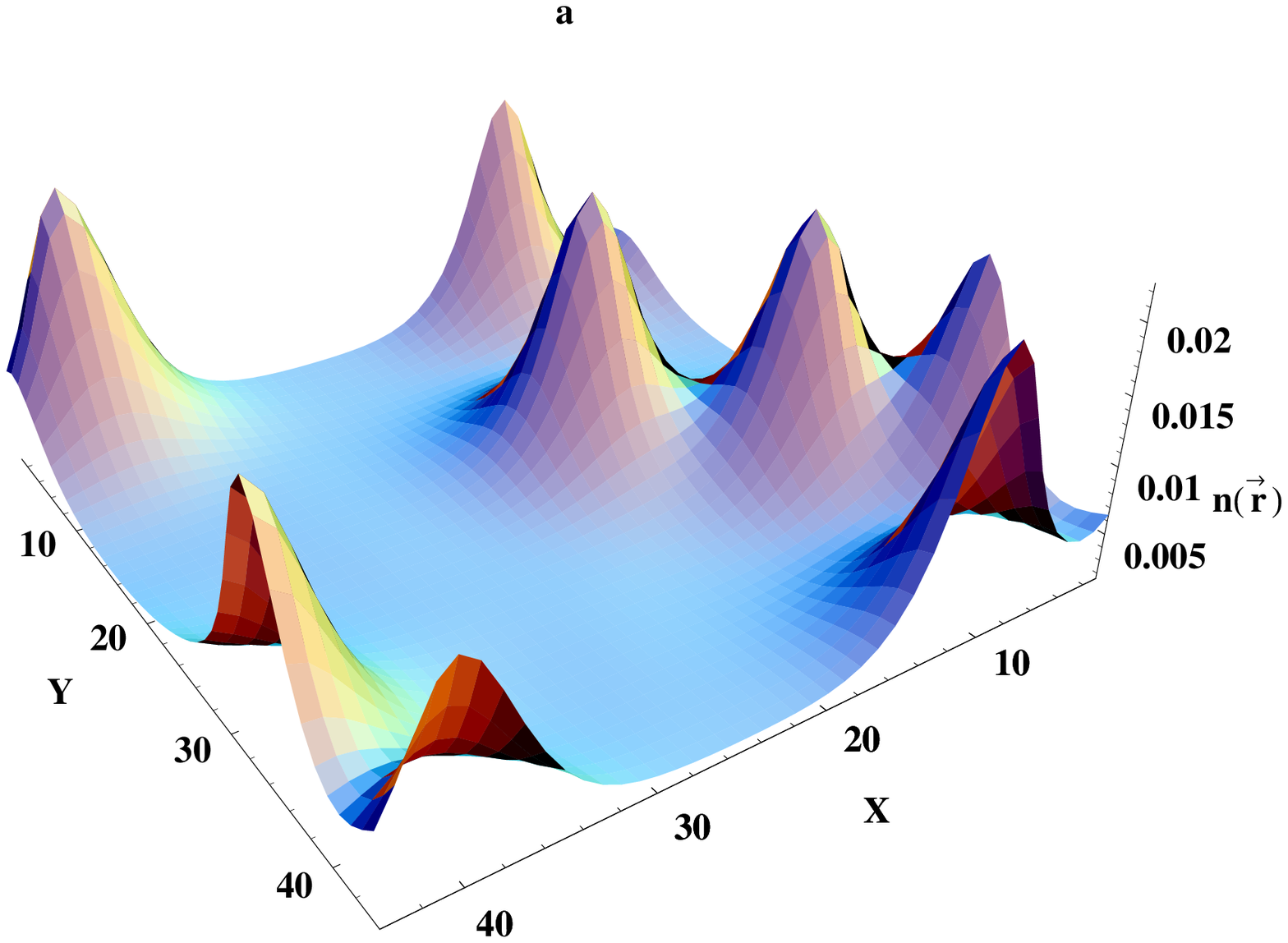}
\includegraphics[width=6 cm]{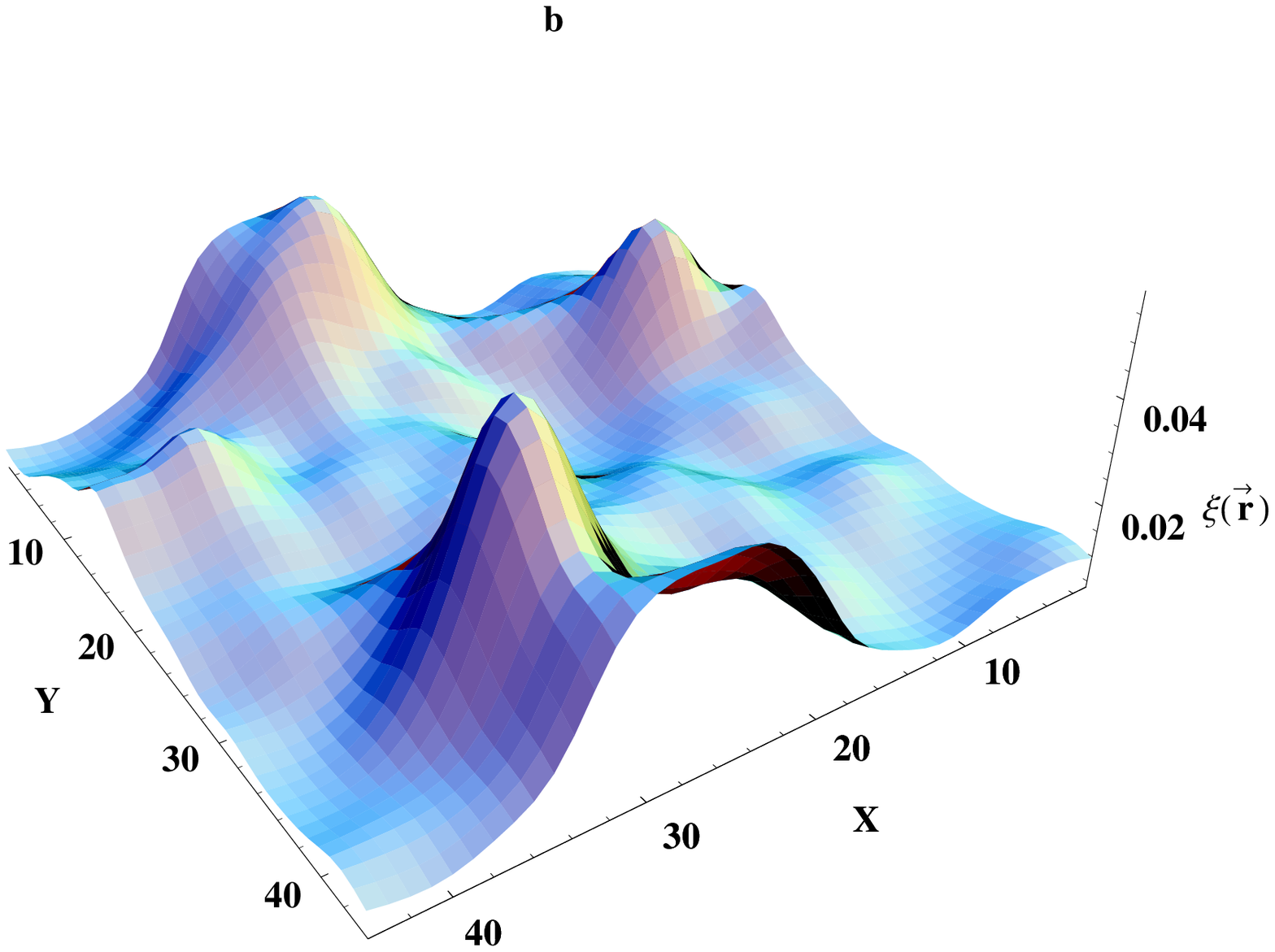}
\includegraphics[width=6 cm]{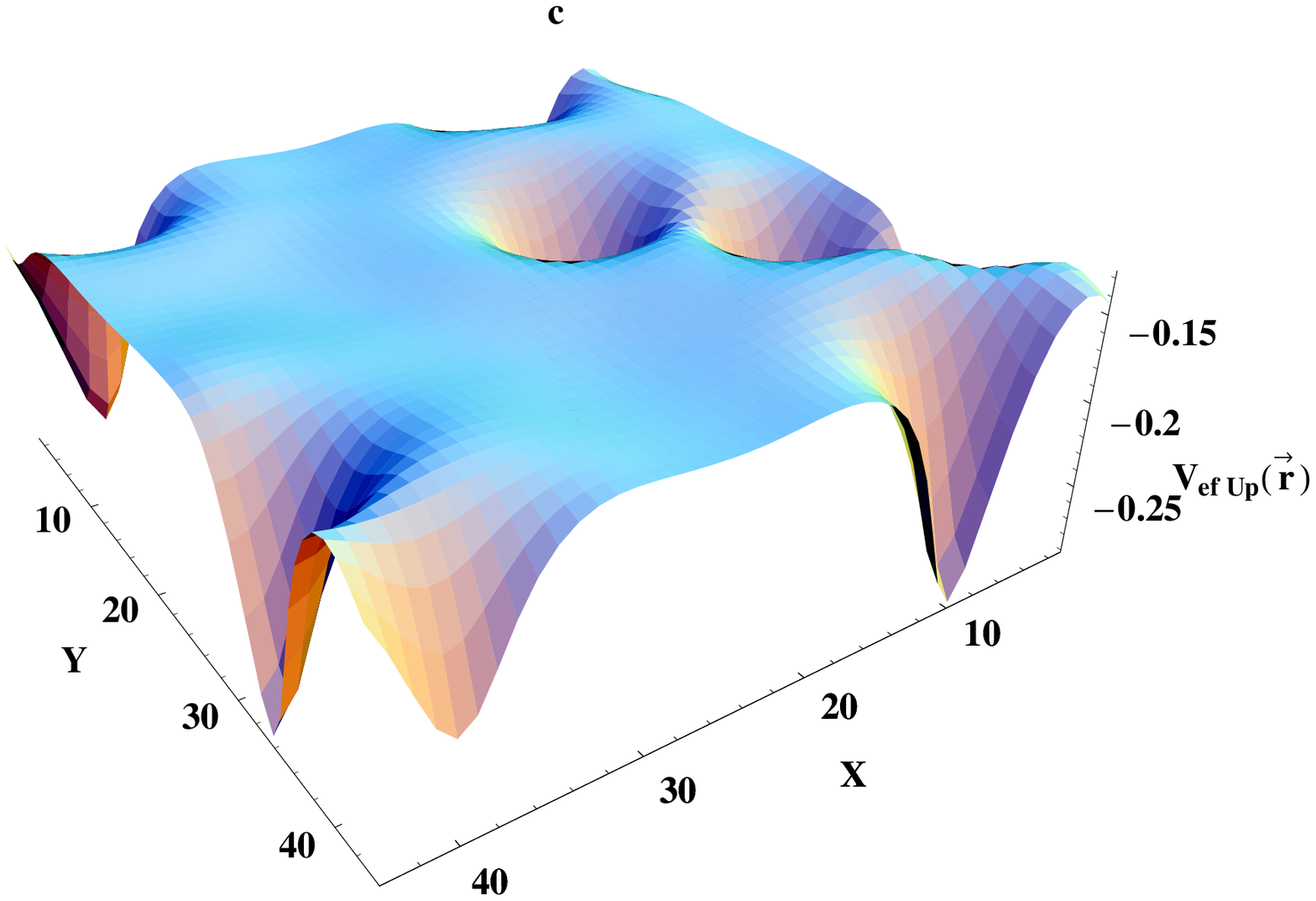}
\caption{(Color online) The density, magnetization per particle $\xi$, and
effective potential distribution at the critical point (minimum of
$\partial\mu /\partial n$), within the TFW theory. The parameters for this figure are as
follows. $d=3$\,a.u., $n_i=2.6\times 10^{-3}$\,a.u., $r_s=8$,
$n=5\times 10^{-3}$\,a.u. One
can see that the magnetization is largest in the low density or
saddle point (for the potential) regions.}
\end{center} 
\label{density}
\end{figure}

To investigate further the effect of the disorder on the spontaneous
magnetization of the electron gas, we calculated its magnetic
susceptibility by applying an in plane magnetic field to the sample
and measure $\chi= M/B_{\parallel}$
at different densities. The effect of the parallel field is only to
create a Zeeman splitting in the energy of electrons but the field
does not affect their orbital motion as the system is purely
two-dimensional.
As can be seen in Fig.\,3, even for low fields, both the
magnetization and the susceptibility reach their maximum at the
density where $\partial \mu /\partial n$ is minimum.
\begin{figure}[h]
\begin{center}
\includegraphics[width=6 cm]{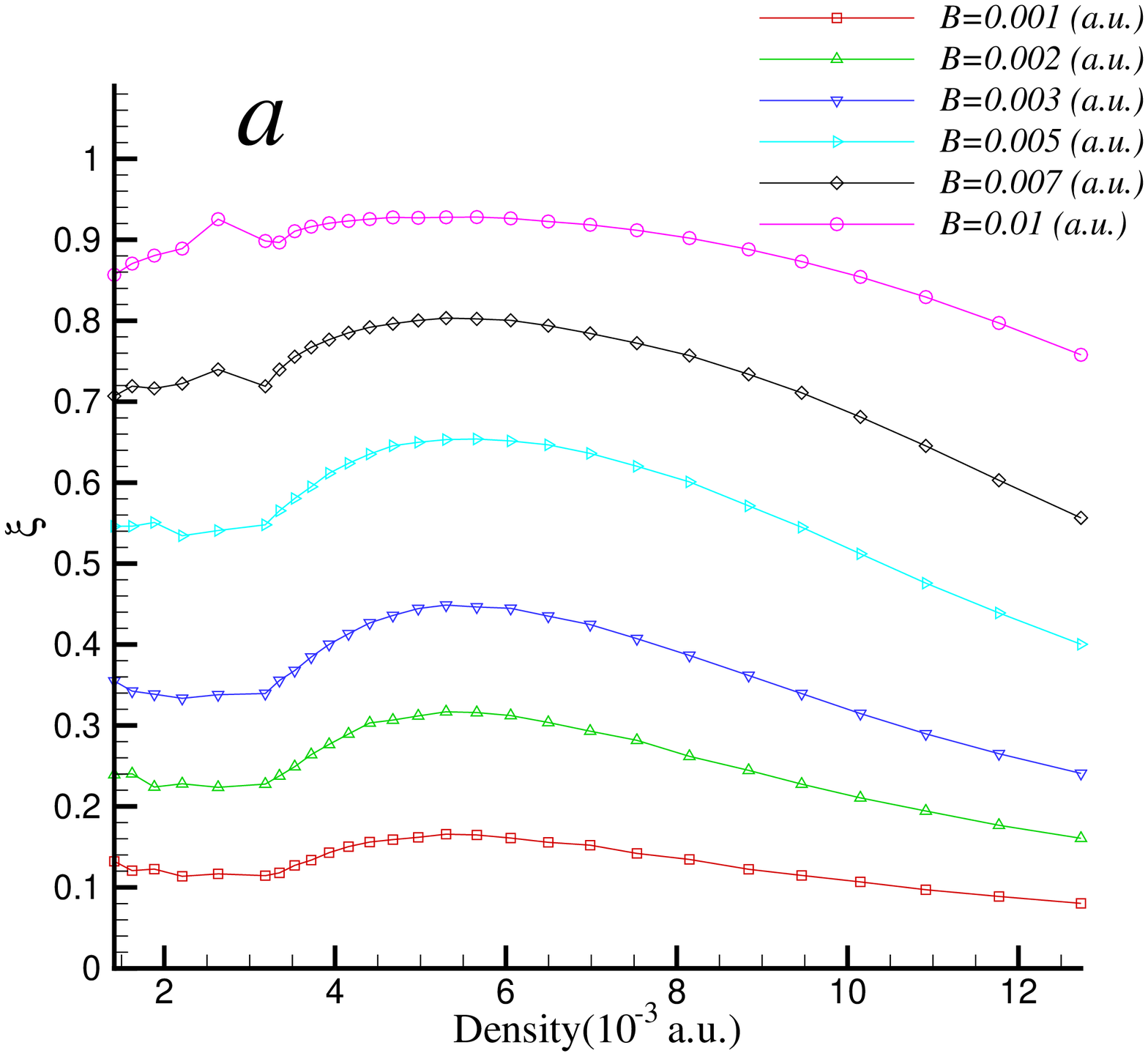}
\includegraphics[width=6 cm]{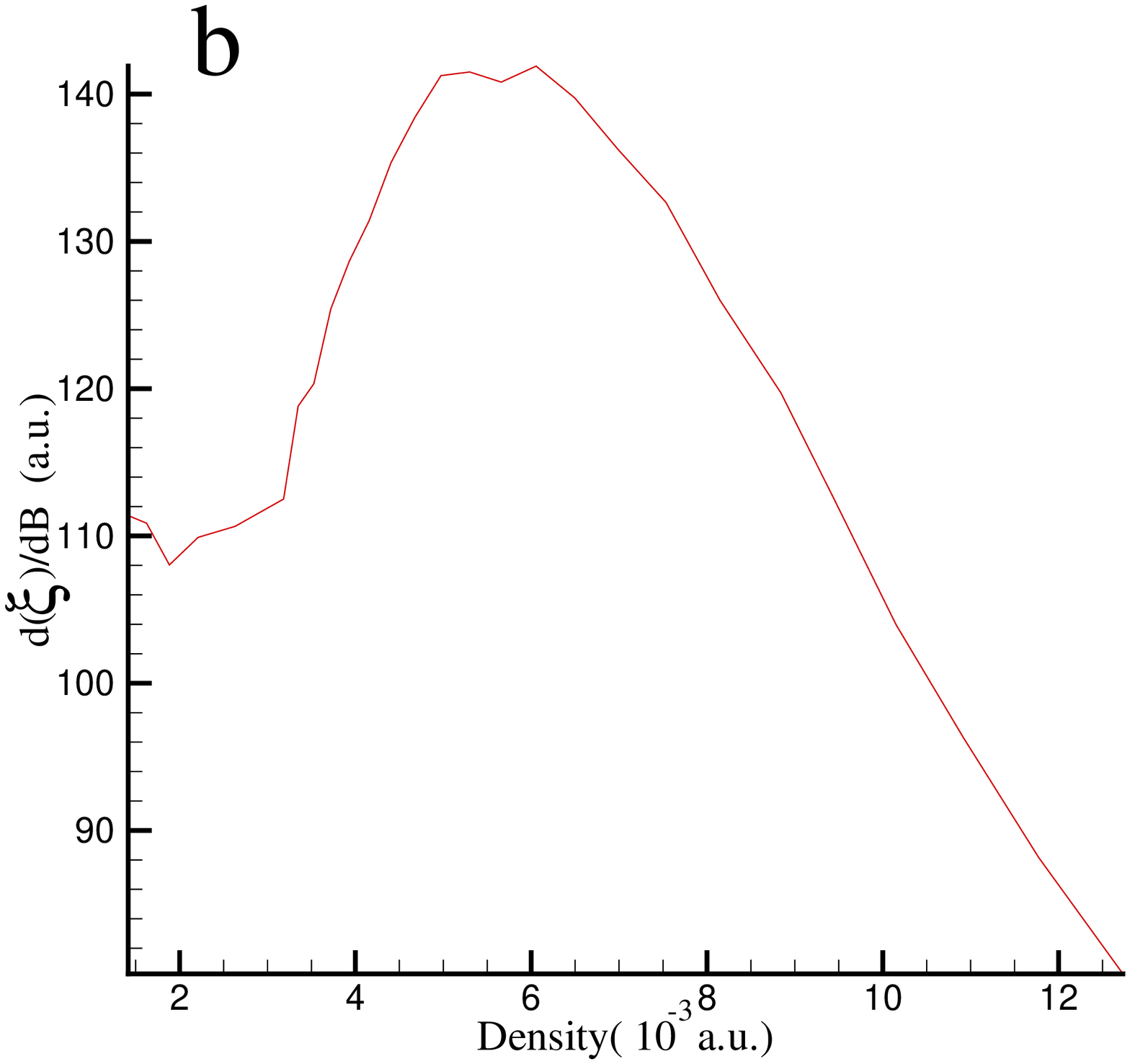}
\caption{(Color online) The spin polarization $\xi$ (a) and its derivative
$\chi$ (b) versus the density for different strengths of the
applied parallel field. Impurity parameters are the same as in
previous figures. It is seen that the maximum in the
magnetization or $\chi$ occurs exactly at the density where
$\partial \mu /\partial n $ was minimum. 
}
\end{center} 
\label{chi}
\end{figure}

At its minimum, we have $\partial \mu /\partial n <0$, and according to
\cite{xie,dultz} we are at the percolation transition.
An added electron will mostly extend in the
percolation region where the density is very low and effective potential flat, and according
to our exchange-correlation (XC) functional at low densities, the
added electron tends to become magnetized. This tendency is strongest
only at the percolation point since at high densities the XC effects
do not favor magnetization, and below percolation densities the
added electron will tend to be localized in the puddles where
again the density is high.
To confirm this picture, we need to know the density distribution
of the highest occupied and lowest unoccupied states, below and
above the Fermi level respectively, where the subtracted or added
electrons localize.
TFW approximation can not solve for energy levels and eigenstates, but one can
extract these levels from the solutions of the KS
equations.
For the same disorder parameters as above, but for a twice coarser
mesh, we have solved the KS
equations and we are displaying in Fig.\,4 the highest
occupied (HOS) and lowest unoccupied (LUS) states at the critical density
where $\partial^2 \mu /\partial n^2 =0$.
\begin{figure}[h]
\begin{center}
\includegraphics[width=6 cm]{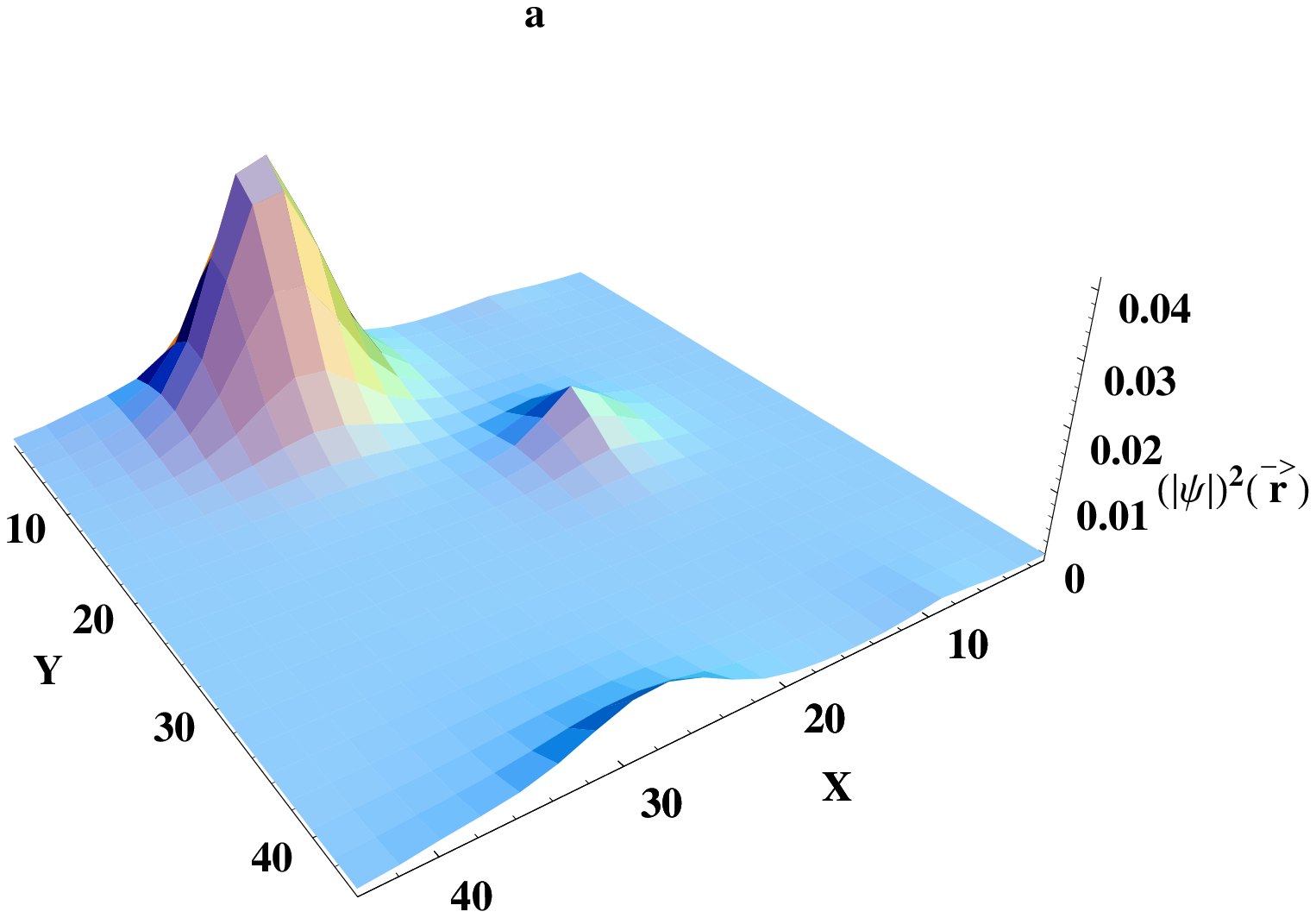}
\includegraphics[width=6 cm]{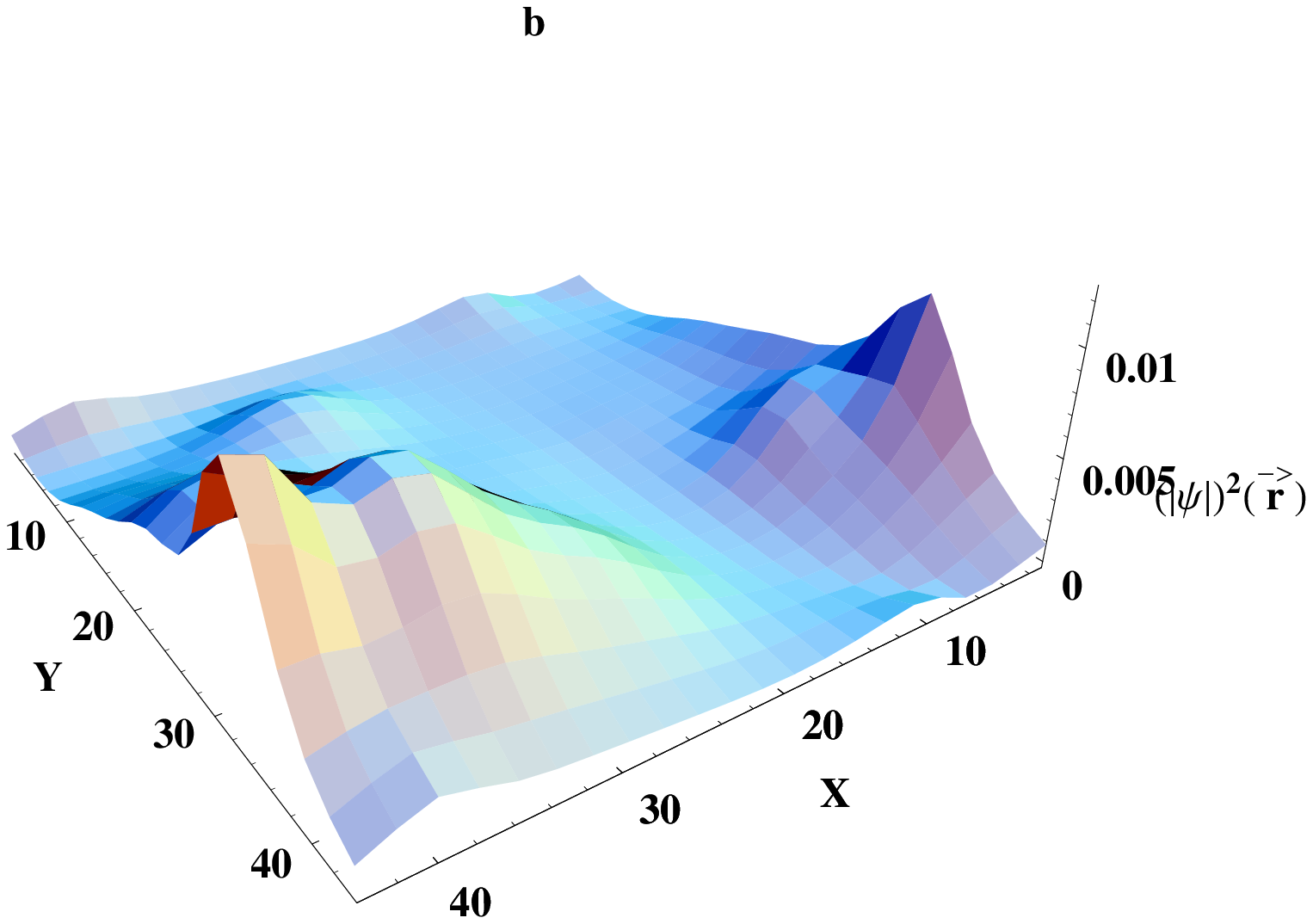}
\caption{(Color online) The density distribution of the highest
occupied, HOS (a) and lowest unoccupied, LUS (b) states just below and
above the Fermi level, respectively, at the critical density where
$\partial \mu /\partial n $ is minimum. Fig. 5 shows that they are both down-spin states.}
\end{center} 
\label{hos-lus}
\end{figure}

To see whether the HOS and LUS are localized or not, one can compute the 
inverse participation ratio defined as,  
${\rm IPR}=\int |\psi|^4 d^2r/\int |\psi|^2 d^2r$.
For an exponentially localized state in 2D, it gives the square of the 
inverse localization length. From the KS equations, we have 
determined the band structure of the finite sample
and for the considered system, it turns out that both the HOS and LUS 
are in the down-spin state, consistent with a large magnetic susceptibility. 
The IPR is plotted as a function of the energy of the states and displayed 
in Fig.\,5.
\begin{figure}
\begin{center}
\includegraphics[width=6 cm]{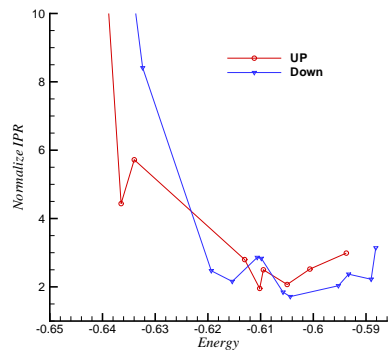}
\caption{(Color online) IPR versus the energy in a.u. for the states around the Fermi level (which is at -0.63 a.u.), at the critical density where $\partial \mu /\partial n $ is minimum. }
\end{center} 
\label{ipr}
\end{figure}
One notices that above the Fermi level, it reaches an almost constant 
value of 3, and below this energy it starts going up implying that the 
HOS seems localized and the LUS is delocalized. This confirms that at 
the chosen density where $\partial \mu /\partial n $ is minimum, occupied 
states are localized and unoccupied ones delocalized, meaning that one 
is also at the MIT point. 



{\it Summary}--In summary, we have investigated the ground state
density and magnetization distribution of an interacting electron
system in 2D in the presence of remote impurities. 
Within the purely thermodynamic calculations based on LSDA, we found that the percolation
transition density where the inverse capacitance $\partial\mu /\partial n$ 
exhibits a minimum, coincides with the MIT based 
on the IPR results. At the same critical density the magnetization and 
magnetic susceptibility display
a maximum where the system goes into a partially polarized
ferromagnetic state. 
and maximum in magnetization are correlated. These calculations suggest that
the percolation transition, the ferromagnetic transition and the
MIT points are all correlated.

This work is supported by TUBITAK and TUBA.
K. E. Would like to thank Prof. A. P. Young and the hospitality of the 
Physics Department of UC Santa Cruz and Bilkent University where part of this work was performed.


\end{document}